\documentstyle[12pt,times,epsf]{article}

\begin{document}

\pagestyle{empty}

{\bf Tribological Behavior of Very Thin Confined Films}\\

\noindent
{\bf M. H. M\"user }\\
Institut f\"ur Physik, WA 331\\
Johannes Gutenberg-Universit\"at Mainz\\
55099 Mainz, Germany\\

\noindent
{\bf ABSTRACT}\\

The tribological properties of two smooth surfaces in the presence of a
thin confined film are investigated with a generic model for the
interaction between two surfaces and with computer simulations. 
It is shown that at large normal contact pressures, an ultra thin film
automatically leads to static friction between two flat surfaces - even if
the surfaces are incommensurate. Commensurability is nevertheless the
key quantity to understand the tribological behavior of the contact.
Qualitative differences between commensurate and incommensurate contacts
remain even in the presence of a thin film. The differences mainly
concern the thermal diffusion of the contact and the transition between
smooth sliding and stick-slip.\\

\noindent
{\bf INTRODUCTION}\\

Understanding the dynamics of a system under shear containing a
confined fluid is intimately connected with understanding
the effects that are
invoked through the corrugation of the confining walls.
In the case of bare, flat surfaces, not only the degree of corrugation
is relevant, but more importantly, the correlation of the corrugation
in the upper wall and in the lower wall, i.e., commensurate systems
exhibit wearless static friction while incommensurate systems do not~[1-3].
This situation is dramatically changed if small amounts
of ``fluid'' are injected into the interface and the fluid atoms
do not form covalent bonds with neither surface~[4].
Independent of the type of
commensurability, static friction can be expected to occur.
Here we will discuss why such mobile atoms in the interface lead
to friction and address the question whether the differences between
commensurate and incommensurate systems are remedied in the presence
of a fluid layer.

The studies discussed here address fundamental issues rather than
questions of direct, practical use. How does a tiny amount of
fluid/contamination between two perfectly flat, crystalline surfaces
alter the tribological behavior of the contact and what are the
implications of commensurability in the presence of a thin film?
Experimentally, it might be impossible to study these effects
satisfactoryly, because clean surfaces are  hard
to obtain even in UHV, e.g., the contaminant may reside within the bulk
and diffuse to the surfaces only after sliding has been
initiated.\\

\noindent
{\bf THEORY AND COMPUTER SIMULATION MODEL}\\

A simple, idealized model to treat interactions between two flat walls
with only atomistic roughness
allows a large number of predictions~[5]. The main feature
of the model is that two surfaces pay a local energy penalty that
increases exponentially fast as the distance between the surfaces is
decreased or the overlap is increased. The consequences of this
model are among others: (i) Commensurate systems show a static
friction coefficient $\mu_s$ that is independent of the 
area of contact $A_c$, (ii) for  amorphous crystalline 
contacts  $\mu_s \propto 1/\sqrt{A_c}$ is found, and (iii) for incommensurate
contacts $\mu_s = 0$. In this case, sliding is only opposed by a viscous
drag force. These predictions are based on the analysis of the
Fourier transform of the surface modulation. $\mu_s$ can only
be independent of $A_c$ if 
the upper and the lower wall's corrugation
is systematically correlated.  Atoms that are injected into
the interface can easily accommodate the surface modulation of both surfaces
simultaneously: The ``fluid'' atoms sit at positions where the spacing
between both walls is maximum. Once the normal pressure becomes large,
the atoms are caught in these position and they can only escape via thermal
activation, which one can assume to take place on long time scales only.
If one tries to initiate relative sliding of the walls,
the available free volume of the atoms will decrease. Thus 
the total energy of the system increases, which results in
a force opposing the initiated motion.

However, there is an important difference between commensurate
and incommensurate walls if thermal activation and diffusion of the
fluid atoms is taken into account. Consider a system that is only
subject to thermal noise. In the commensurate case, fluid atoms
jump into (more or less) equivalent positions. Therefore, the relative
lateral equilibrium position of the top wall does not change even if the
fluid atoms undergo diffusion. In other words, there is a well-defined
free energy profile of the contact that has the periodicity
of the system. Therefore, static friction is an equilibrium phenomenon.
In the incommensurate case, the expected
situation is strikingly different. The fluid atoms jump  into inequivalent
positions as they undergo thermally activated diffusion. With each such
jump, the lateral, relative equilibrium position of the walls shifts 
slightly. If we allow the fluid to explore the whole phase space,
all relative, lateral positions of the two walls are identical.
Therefore the free energy is not a function of the relative, lateral
displacement of the two walls unlike the commensurate case. For
incommensurate systems, static friction is a non-equilibrium phenomenon.
From this discussion, one would expect an exponential slowing
down of thermal diffusion with contact area of the whole contact in the 
commensurate case (at fixed normal pressure), while creep motion would
be expected for incommensurate surfaces.

A generic model is emploied in order to analyze the thermal motion of
a mechanical contact including a thin confined layer of atoms.
It consists of two (111) surfaces whose atoms are harmonically pinned
to their ideal lattice sites. Periodic boundary conditions are emploied
in the plane of the two walls. The fluid atoms, which are confined between
the walls, interact with Lennard Jones potentials among each other
and with the walls. For further details of the model, we refer to
Refs.~[3-5].\\

\noindent
{\bf RESULTS}\\

We first want to report results for the thermal diffusion of the contact.
Commensurate surfaces show the expected exponentially slowing down
with increasing area of contact $A_c$ for a given normal pressure 
$p_\perp$ and temperature $T$.
Of course, a similar slowing down of the diffusion is achieved by
either decreasing $T$ at fixed ($A_c,\, p_\perp$) or by increasing
$p_\perp$ at fixed ($A_c,\, T$)~[3]. Surprisingly, dramatic size effects
are also found for incommensurate surfaces as shown in figure~1.

\begin{figure}[tbhp]
\begin{center}
\leavevmode
\hbox{\epsfxsize=90mm \epsfbox{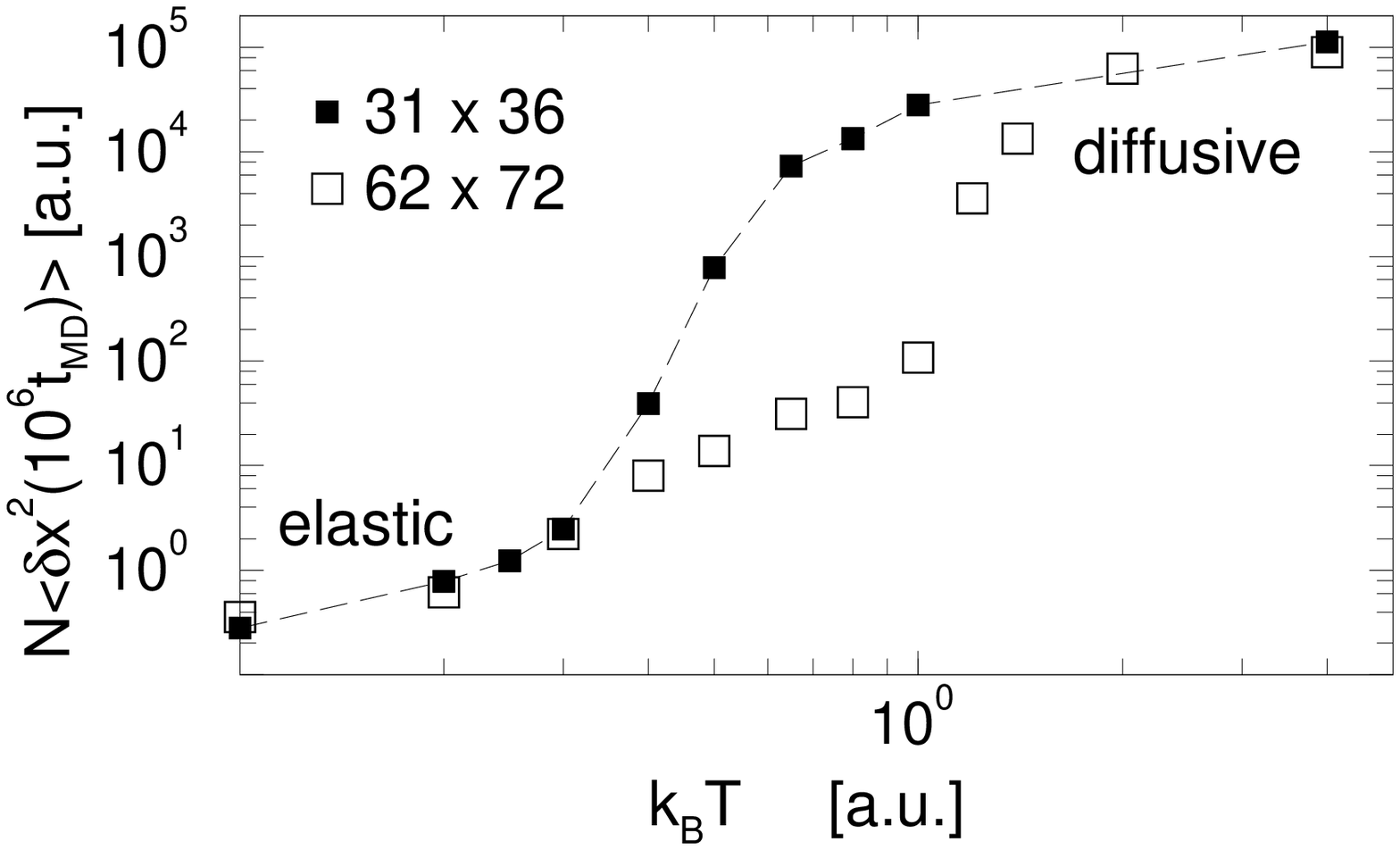}} 
\end{center}
{\small {\bf Figure 1.}
{\it Mean square displacement of incommensurate top wall times number of
 atoms $N$ per surface layer after 10$^6$ MD steps for two 
   system sizes, $N = 31\times 36$ and $N = 62\times 72$,
  as function of temperature.} }
\end{figure}

At small temperatures, the system appears to be pinned elastically
during a time window of $10^6$ molecular dynamics (MD) time steps.
The effective lateral stiffness increases linearly with the area of
contact $A_c$ similar to the net friction force in Amontons law,
which is proportional to $p_\perp \times A_c$. This generalized
Amontons law for elastic pinning has also been observed 
experimentally~[6].
It can be extracted from the small magnitude of the MSD and the
decrease of the MSD with the number of atoms $N$ per surface layer
(in Lennard Jones units $N$ and $A_c$ are related by a factor close to
unity). At large temperatures, the contact can be considered ``diffusive''.
In this regime the same scaling factor, namely $N$, collapses the data for
the two system sizes. However, in between these two regimes, a non-trivial
size dependence is observed, where the increase of the system size by a 
factor of four decreases the MSD of the top wall by more than a
factor of hundred. It is important to note that the diffusion of
individual fluid atoms is not affected by the size of the upper wall.
In all cases, the MSD of individual atoms is larger than $N$ times the
MSD of the top wall.

This observation goes beyond the theoretical considerations
from the last section. The pinning mechanism seems to be even more
effective than anticipated. The large wall still appears to be pinned
even when individual fluid atoms have typically diffused over significantly
more than hundred lattice constants. We speculate that the indirect
long-range interactions between monomers mediated through the top 
wall's center-of-mass may be the reason for the unexpected slowing down
of the equilibration of the top wall. (Remember that all lateral position of
an incommensurate top wall are equivalent provided that the fluid is in
thermal equilibrium.) Long-range interactions have been
shown to frequently slow down equilibration~[7].

There is nevertheless a significant difference at what system size
commensurate or incommensurate systems appear to be pinned at 
fixed  ($T,\, p_\perp$). Other parameters like coverage of the fluid,
the strength of the repulsive interactions, etc.
play a much less crucial role. In our model, commensurate
systems always pin at smaller system sizes as compared to
incommensurate systems. This is also reflected in the
static friction, which is shown in figure~2.
\begin{figure}[tbhp]
\begin{center}
\leavevmode
\hbox{\epsfxsize=80mm \epsfbox{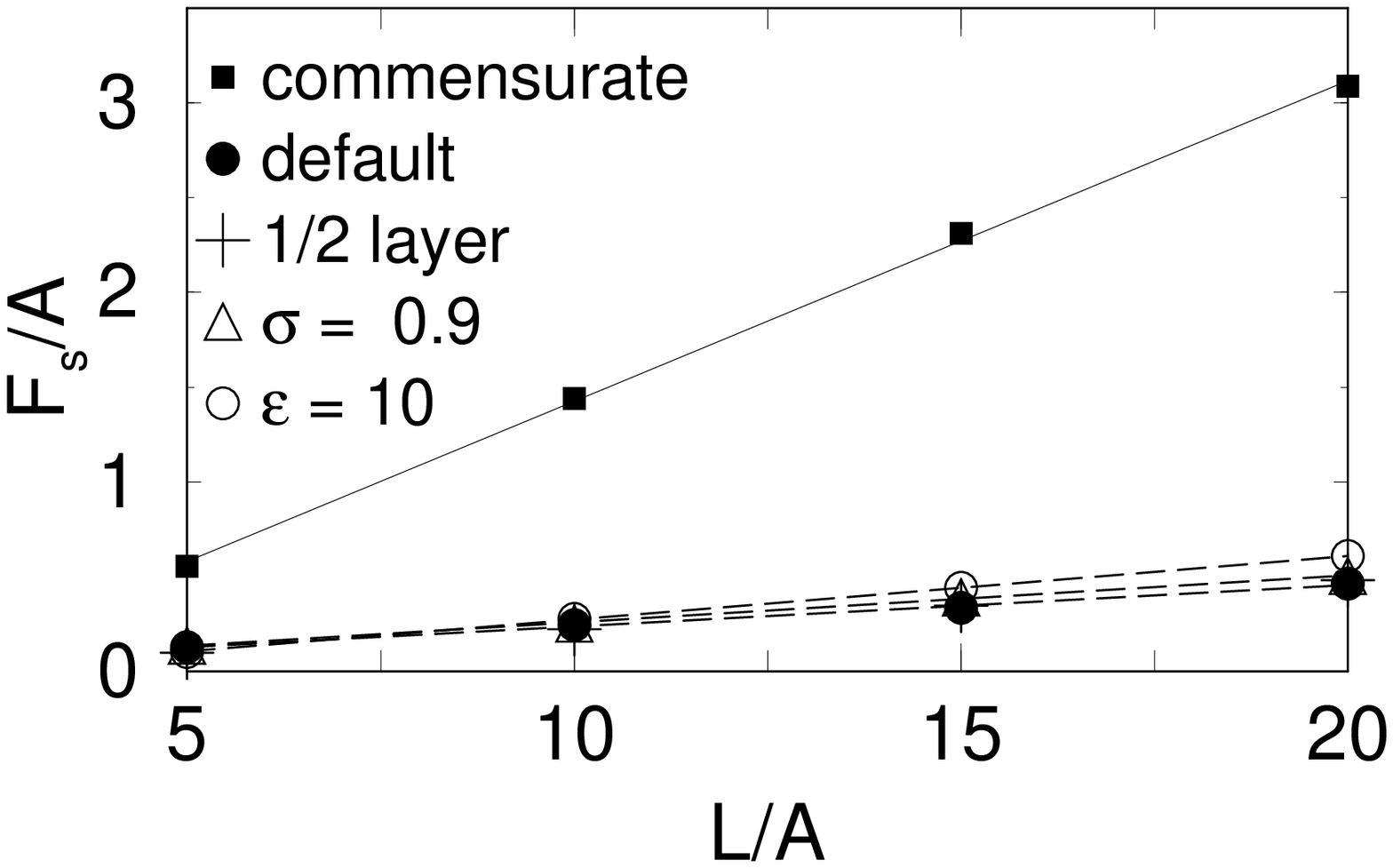}} 
\end{center}
{\small {\bf Figure 2.}
{\it Shear pressure $p_s = F_s/A$ versus 
normal pressure $p_\perp = L/A$ for different model systems.
The default system consists of two flat incommensurate walls with
a quarter layer of fluid confined in between them. The default
Lennard Jones interaction parameters are
$\sigma = 1$ and $\epsilon = 1$.  } }
\end{figure}
The default system in figure~2 consists of a quarter layer lubricant
composed by simple Lennard Jones atoms that interact with
$V(r) = 4\epsilon[(\sigma/r)^{12}-(\sigma/r)^6)]$ with the
default values for $\sigma$ and $\epsilon$ being unity. Potentials
are cut-off such that interactions are purely repulsive. In the absence
of free surfaces, the main effect
of the long-range adhesive part can be absorbed into the load.
We note in passing that the net interaction between two
fluid atoms is nevertheless attractive due to their elastic interactions
with the confining walls. As long as the two confining walls are
incommensurate, the static friction coefficient $\mu_s$, 
which can be defined as the
slope of the curves in figure~2, is relatively independent of the
model: The inequality $0.0022 < \mu_s < 0.0035$ holds for the
default system, and systems for which one parameter is changed, namely,
a half layer lubrication instead of a quarter lubrication, 
$\epsilon = 10$, and $\sigma = 0.9$. Rotating the walls to make them
orientationally perfectly aligned, however, increases $\mu_s$
by nearly a factor of five
to $\mu_s  \approx 0.17$.

While these differences in $\mu_s$ are merely quantitative, there is a
qualitative difference in the transition from stick-slip to smooth sliding.
Here, we calculate the (average) kinetic friction coefficient
$\mu_k = F_k/L$ 
for a system that is pulled with a spring of stiffness $k$. For small values
of $k$ the system shows stick-slip for large values of $k$ smooth sliding
at constant velocity is found. The average friction force is shown in
figure~3.
The commensurate surfaces show similar behavior as 
dry commensurate surfaces. For a weak spring, the frictional forces are 
large. In this regime, the top wall shows large scale stick-slip motion.
In an intermediate regime, in which atomic scale ratcheting is found,
the frictional forces decrease dramatically upon increasing $k$. 
\begin{figure}[h]
\begin{center}
\leavevmode
\hbox{\epsfxsize=90mm \epsfbox{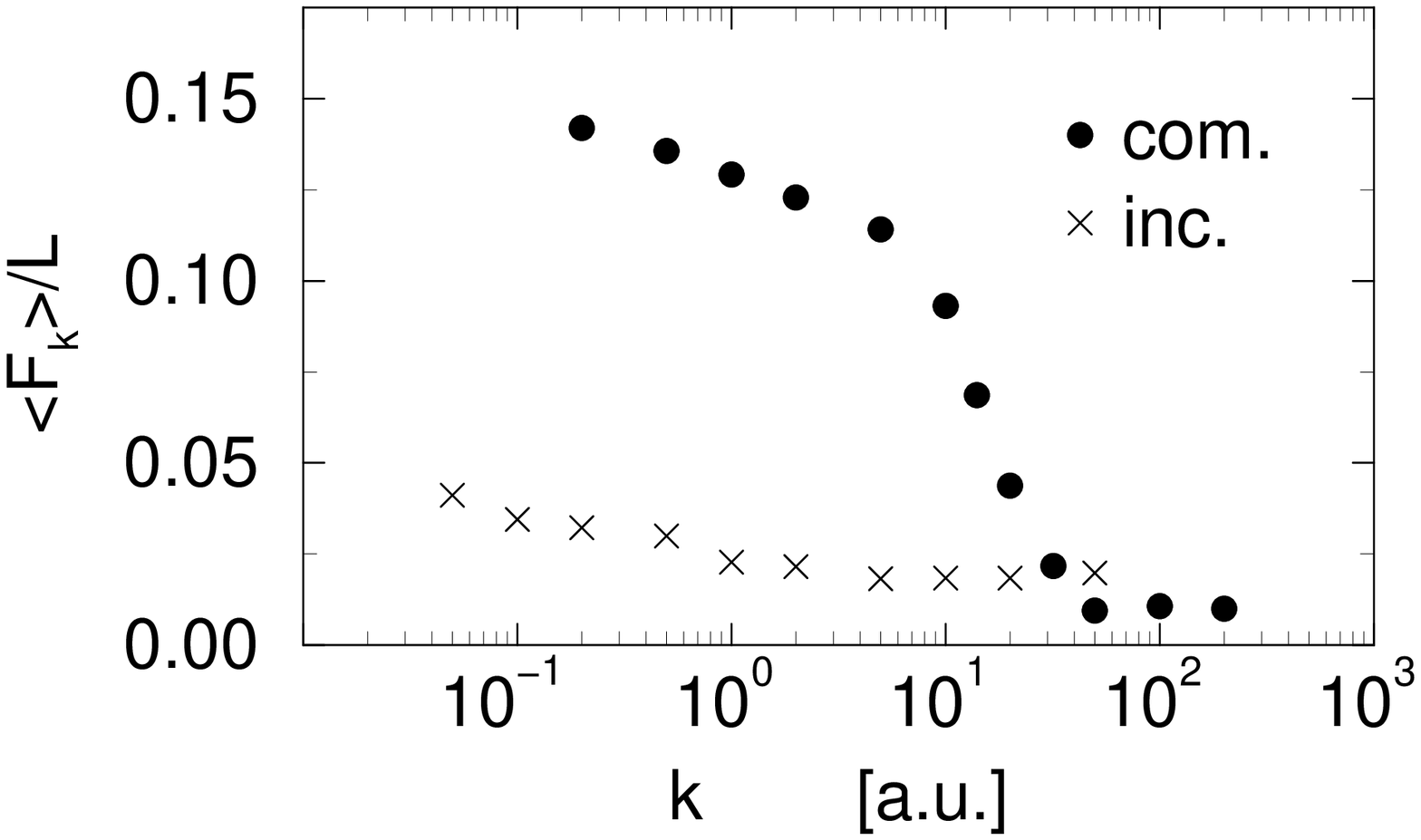}} 
\end{center}
{\small {\bf Figure 3.}
{\it  Average kinetic friction force per load as a function of the stiffness
 $k$ of the spring that drives the system for commensurate and incommensurate
 surfaces separated by a quarter layer of fluid atoms.
  } }
\end{figure}
\newpage

At large $k$, the upper wall
goes up and down the free energy surface adiabatically, resulting in small
frictional forces.
 Of course, in the smooth sliding regime, $F$ vanishes
all together in the limit $k \to \infty$ and driving velocity $v \to 0$.
The incommensurate surfaces, however, can not be driven adiabatically
resulting in considerably larger friction at large values of $k$.
We want to note in passing that commensurate surfaces show periodic
stick-slip events while the incommensurate surfaces with boundary lubrication
show much more erratic stick-slip events.

So far, we have only studied the friction between flat surfaces.
However, boundary lubrication also effects the friction between curved
surfaces. In order to study effects related to the curvature of surfaces,
we have studied a Hertzian contact of a curved tip  on a flat surface
as a function of commensurability and boundary lubrication~[8].
Due to the finiteness of the contact, even incommensurate contacts
show friction, although it can certainly be considered insignificantly
small - as long as the friction remains wear free. The situation changes
when a thin layer with mobile atoms is present on at least one of the
two surfaces. In the study of such systems, the interaction of the tip
with the lower wall and with the mobile atoms is purely repulsive.
This choice eliminates adhesive effects such as the so called offset
load. All other interactions are chosen to be attractive.
Some representative results are shown in figure~4.

Non-adhesive commensurate tips show the expected linear relationship between
friction force and load, e.g., 
$F_s = L^\alpha$ with $\alpha = 1$. 
A dry amorphous tip on a crystalline substrate shows sublinear
behavior with $\alpha = 2/3$, which agrees well with friction force microscopy
experiments~[9]. 
Dry incommensurate surfaces show the smallest friction
forces with an insignificant increase in $F_s$ force with
increasing $L$. As soon as a thin film is present in the interface, 
the wearless friction increases dramatically. However, unlike for the
friction between flat surfaces, sublinear behavior is found. In the present
study, the exponent $\alpha = 0.85$ is found, which we do not believe to be
universal. Details of the friction-load dependences will depend
among other on the wetting properties of the fluid.

\begin{figure}[h]
\begin{center}
\leavevmode
\hbox{\epsfxsize=90mm \epsfbox{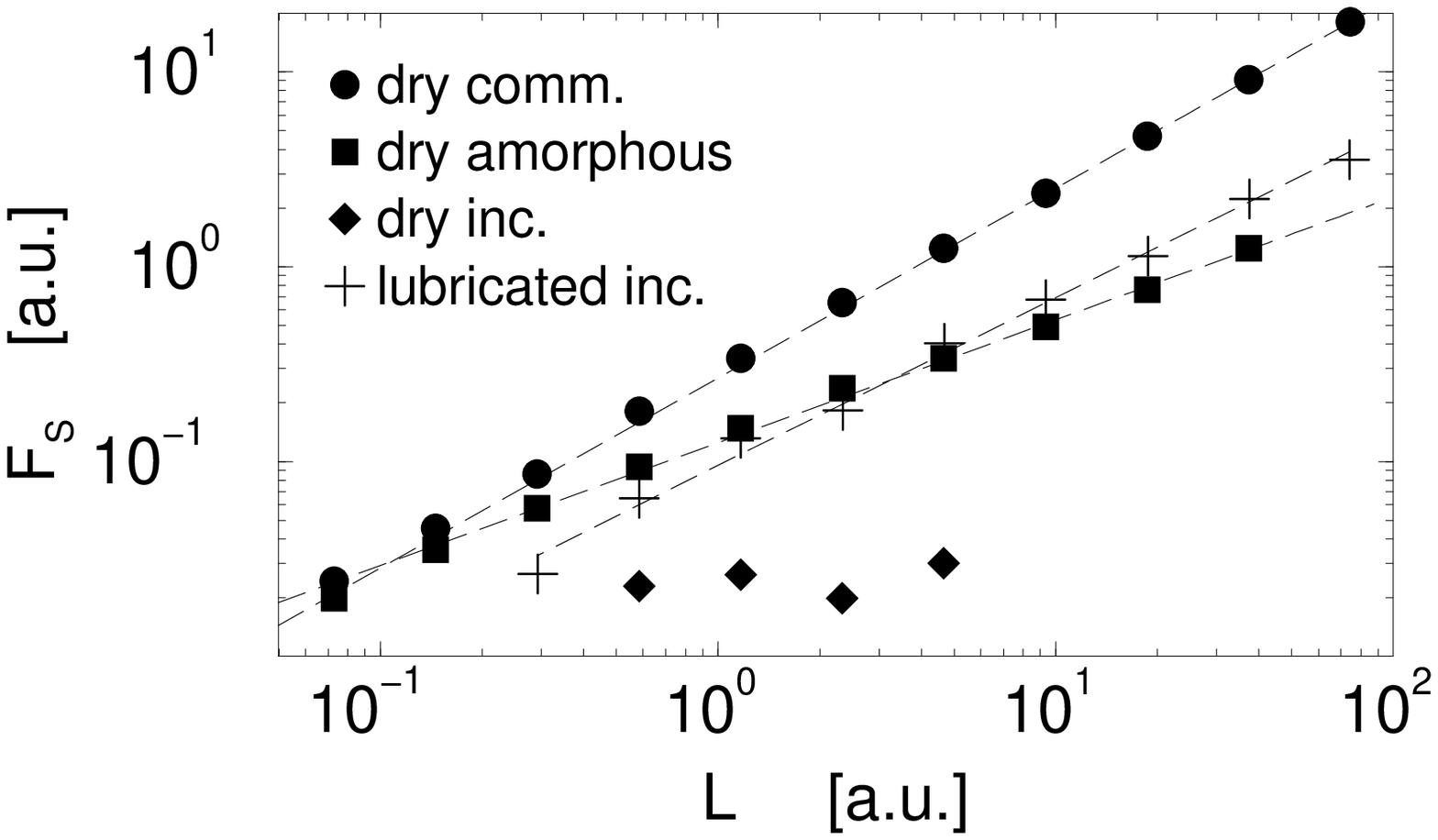}} 
\end{center}
{\small {\bf Figure 4.}
{\it  Static friction force $F_s$ vs load $L$ for different curved tips on 
      a crystalline substrate.
  } }
\end{figure}

\newpage

\noindent
{\bf CONCLUSIONS}\\

We have shown that the commensurability of two walls confining a thin
fluid film  has systematic implications for the tribological properties
despite the presence of the confined film.  
The differences are both quantitative and qualitative.
Static friction is considerably larger between two flat
commensurate surfaces than between incommensurate surfaces.
In the case of a Hertzian contact, 
the friction-force load dependence can even be qualitatively
different, e.g., incommensurate contacts may have a sublinear dependence
on the load. The most striking difference, however, is the transition
from stick-slip to smooth sliding. In our studies, commensurate
surfaces show a dramatic decrease in kinetic friction from
stick-slip to smooth sliding. A rather abrupt decrease in kinetic friction
is seen as the slipped distances become small. This abrupt
decrease in kinetic friction, which was achieved by increasing the
stiffness of the driving device, is absent for incommensurate
surfaces. It would be an interesting question, how friction control
mechanisms such as proposed in Ref.~[10] would be effected by the
incommensurability of the surfaces in the presence of a thin confined
film.
\\

\noindent
{\bf ACKNOWLEDGMENTS}\\

The author would like to thank K. Binder, M. O. Robbins, and L. Wenning
for discussions. Research supported by the 
Israeli-German D.I.P.-Project No 352-101.\\

\noindent
{\bf REFERENCES}

\noindent
\begin{enumerate}
\itemsep=-2pt

\item M. Hirano and K. Shinjo, Phys. Rev. B {\bf 41}, 11837 (1990);
      Wear {\bf 168}, 121 (1993).
\item M. R. S$\o$rensen, K. W. Jacobsen, and P. Stoltze,
      Phys Rev. B {\bf 53}, 2101 (1996). 
\item M. H. M\"user and M. O. Robbins, Phys. Rev. B {\bf 64}, 2335 (2000).
\item G. He, M. H. M\"user, and M. O. Robbins, Science {\bf 284}, 1650 (1999).
\item M. H. M\"user, L. Wenning, and M. O. Robbins,
      cond-mat/0004494.
\item P. Berthoud and T. Baumberger, Proc. Roy. Soc. Lond. {\bf A 454}, 1615
      (1998).
\item C. Tsallis, cond-mat/0011022; cond-mat/9903356.
\item L. Wenning and M. H. M\"user, cond-mat/0010396 (submitted to 
      Europhys. Lett.).
\item U.D. Schwarz , O. Zw\"orner, P. K\"oster, and R. Wiesendanger,
      Phys. Rev. B {\bf 56}, 6997 (1997).
\item V. Zaloj, M. Urnakh, and J. Klafter, Phys. Rev. Lett. {\bf 82}, 4823 
      (1999).
    
\end{enumerate}

\newpage

\noindent
{\bf APPENDIX}\\

Some additional results are presented on the cond-mat server,
which will not be published in the MRS proceedings 
{\it Dynamics in Confining Systems VI}. The figures presented
here serve to shed further light on the discussions related to
figure 3 and figure 4 in the main manuscript.

The time dependence of the frictional force leading to figure~3 are shown
in figure~5. It is interesting to note the difference in the friction
forces in the commensurate case between atomic ratcheting, where
$F$ is nearly always positive and smooth sliding where the friction force
oscillates. In the incommensurate case, the fluctuations of the friction force
is much smaller for small values of $k$ than for commensurate surfaces.
\begin{figure}[h]
\begin{center}
\leavevmode
\hbox{\epsfxsize=60mm \epsfbox{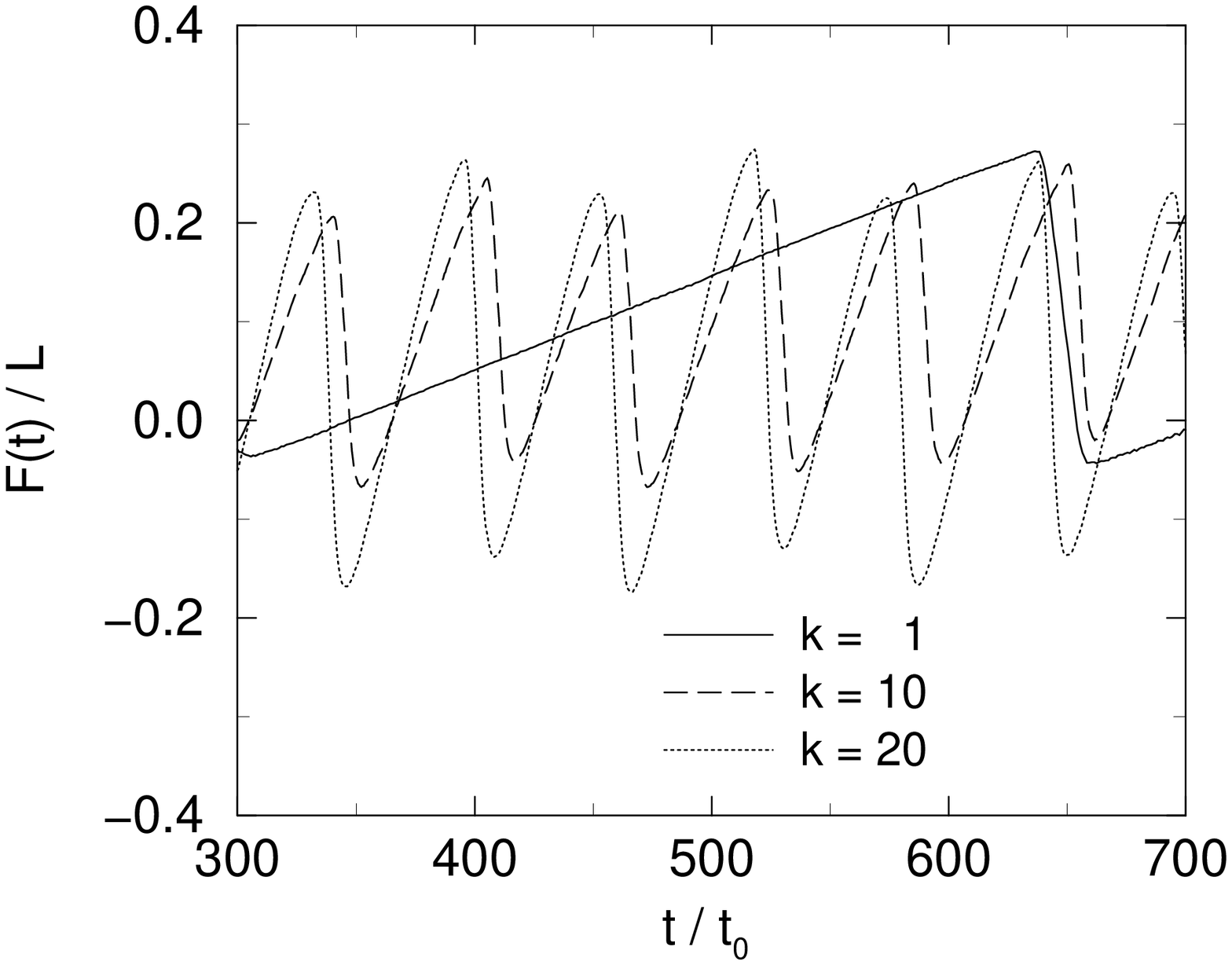}} 
\hbox{\epsfxsize=60mm \epsfbox{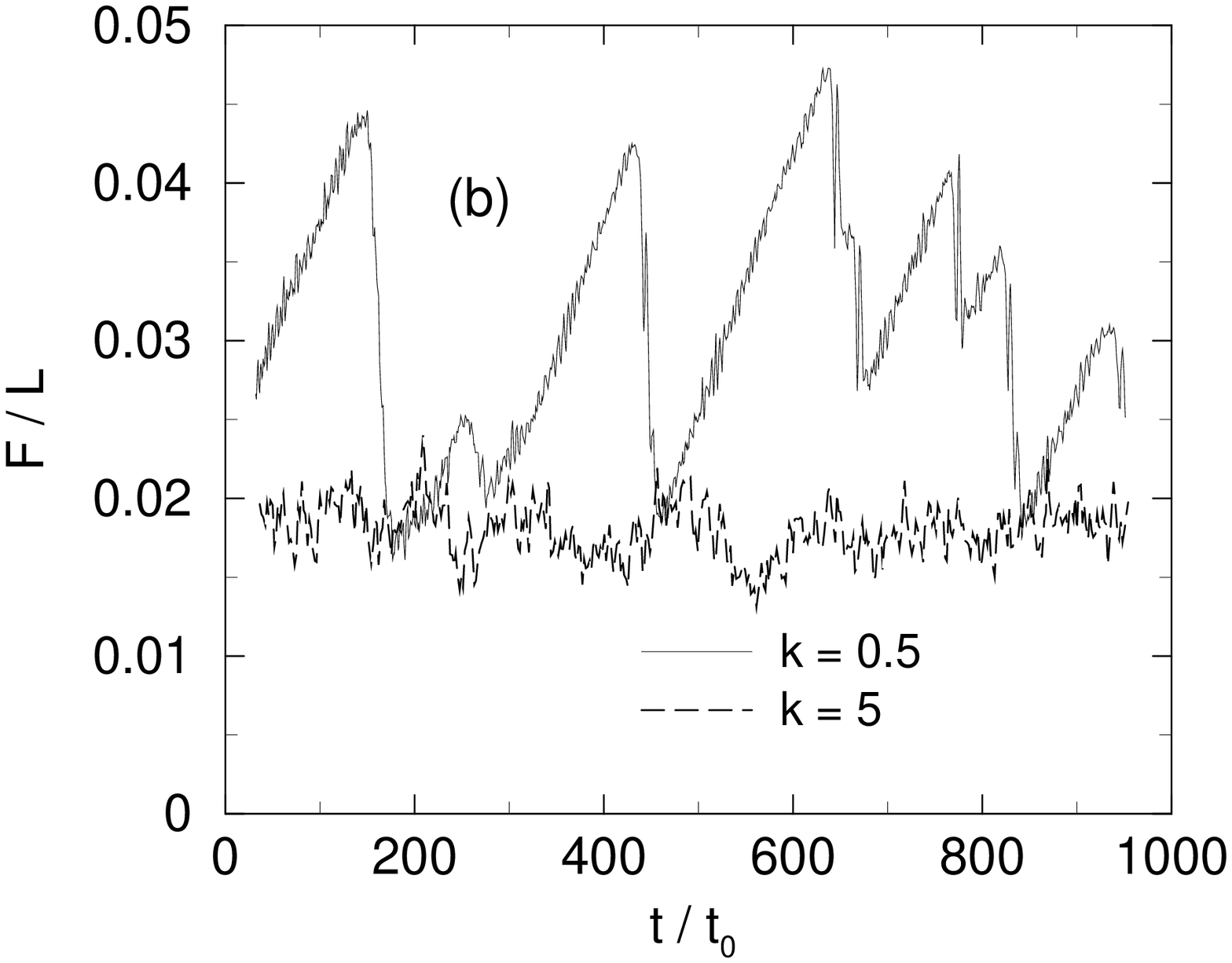}} 
\end{center}
{\small {\bf Figure 5.}
{\it  Time evolution of friction force for different stiffnesses of
      pulling spring. {\bf Left.} Commensurate surfaces.
      {\bf Right.} Incommensurate surfaces.
  } }
\end{figure}

In order to understand the sublinear behavior of the friction load
law shown in figure~4, it is instructive to analyze the normal and
lateral forces acting on individual tip atoms. This is done in 
figure~6. The center of the tip is in direct contact with the substrate.
It carries a considerable amount of load, but the lateral forces
are small and in addition of random sign. At positions where one fluid layer
is found, tip atoms also carry load. This region contributes the
largest fraction to the net friction force, in particular at the entrance
of the tip. The friction mechanism of the confined fluid is similar 
to the one found between flat surfaces. The net friction force of the
tip results from the wetting and squeezing out properties of the lubricant.
\begin{figure}[h]
\begin{center}
\leavevmode
\hbox{\epsfxsize=45mm \epsfbox{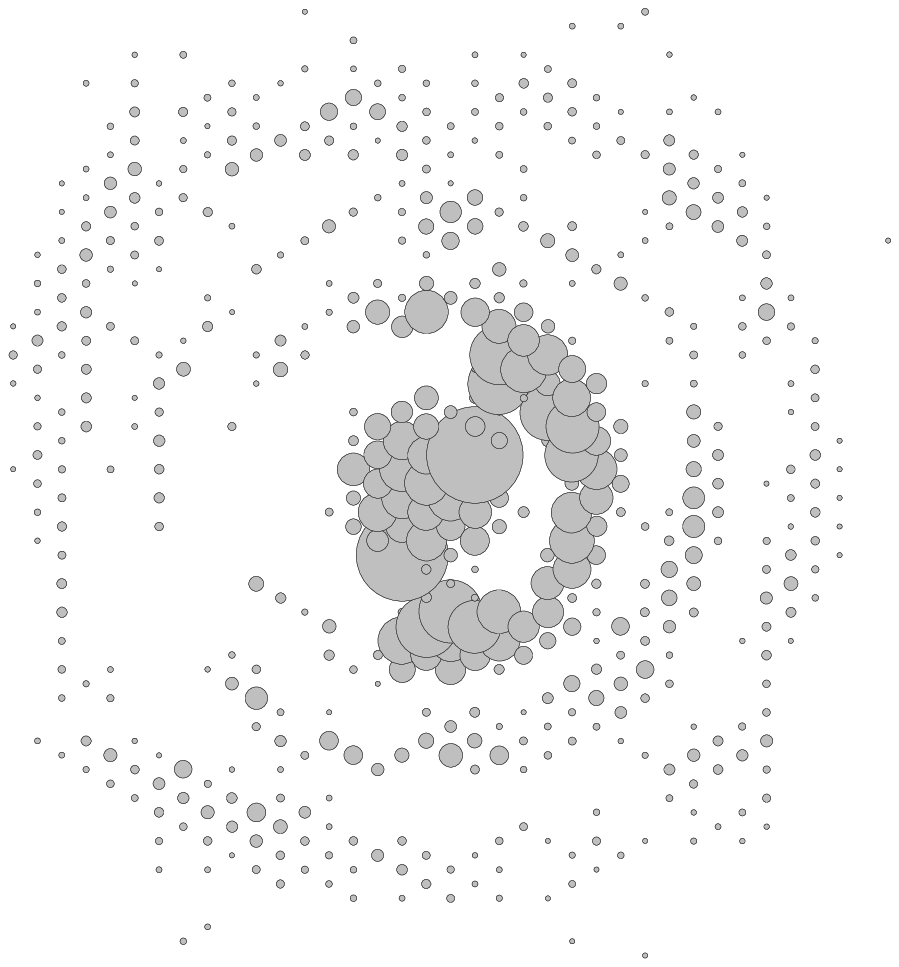}} 
\hbox{\epsfxsize=45mm \epsfbox{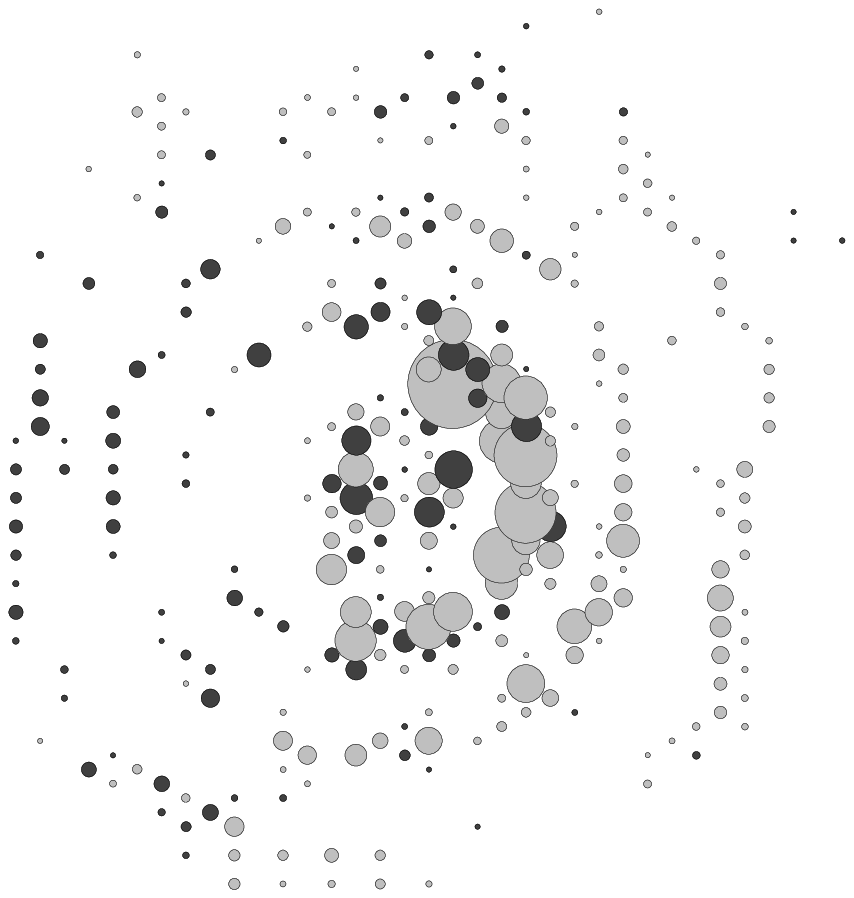}} 
\vspace*{-2cm}
\begin{minipage}{6.5cm} \vspace*{-4cm}

{\small {\bf Figure 6.}
{\it   Force on individual tip atoms.
{\bf Left.} Normal force. {\bf Right.} Lateral force. Light atoms
 represent ``friction'' while dark atoms represent ``antifriction''.
  } }
\end{minipage} 
\end{center}
\end{figure}

\end{document}